\documentclass[preprintnumbers,amsmath,amssymb,floatfix,11pt,prd,onecolumn,showpacs,superscriptaddress,nofootinbib]{revtex4}
\usepackage{graphicx}
\usepackage{epsfig}
\usepackage{bm}
\usepackage{amsfonts}

\begin{document}

\title{ Stability of a non-minimally conformally coupled scalar field in F(T ) cosmology}

\author{\textbf{ Mubasher Jamil}} \email{mjamil@camp.nust.edu.pk}
\affiliation{Center for Advanced Mathematics and Physics (CAMP),\\
National University of Sciences and Technology (NUST), H-12,
Islamabad, Pakistan}
\affiliation{Eurasian International Center
for Theoretical Physics,  L.N. Gumilyov Eurasian National University, Astana
010008, Kazakhstan}
\author{\textbf{ D. Momeni}}
\email{d.momeni@yahoo.com }
 \affiliation{Eurasian International Center
for Theoretical Physics,  L.N. Gumilyov Eurasian National University, Astana
010008, Kazakhstan}
\author{\textbf{ Ratbay Myrzakulov}}
\email{rmyrzakulov@gmail.com}\affiliation{Eurasian International Center
for Theoretical Physics,  L.N. Gumilyov Eurasian National University, Astana
010008, Kazakhstan}

\begin{abstract}
{\bf Abstract:} In this paper, we introduce a non-minimally
conformally coupled scalar field and dark matter in $F(T)$ cosmology
 and study their dynamics. We investigate the stability and phase
  space behavior of the parameters of the scalar field by choosing
   an exponential potential and cosmologically viable form of $F(T)$.
    We found that the dynamical system of equations admit two unstable
     critical points, thus no attractor solutions exist in this cosmology.
      Furthermore taking into account the scalar field mimicking as
       quintessence and phantom energy, we discuss the corresponding
       cosmic evolution for both small and large times. We
       investigate the cosmological implications of the model via
    equation of state and deceleration parameters of our model
        and show that the late time Universe will be dominated by phantom
         energy and moreover phantom crossing is possible.
          Our results have no explicit predictions for inflation and early Universe era. \\

\textbf{Keywords:} Cosmology; torsion; stability; scalar fields; quintessence; phantom energy.

\end{abstract}

\pacs{04.20.Fy; 04.50.+h; 98.80.-k} \maketitle

\newpage
\section{Introduction}

Astrophysicists are convinced that the observable
universe is in a phase of rapid accelerated expansion and commonly termed it `dark energy' (DE) possessing negative pressure and
positive energy density. This conclusion has
been supported by several astrophysical data
findings of supernovae SNe Ia {\cite{c1}}, cosmic microwave
background radiations via WMAP {\cite{c2}}, galaxy redshift surveys
via SDSS {\cite{c3}} and galactic X-ray {\cite{c4}}.
 Although the phenomenon of dark energy in
cosmic history is very recent $z\sim0.7$, it has opened new areas in
cosmology research. The most elegant and simple resolution to DE is the
cosmological constant \cite{c7} but it cannot resolve fine tunning and cosmic coincidence
problem. Hence theorists looked for other
alternative models by considering the dynamic nature of dark energy
like quintessence scalar field \cite{quint}, a phantom energy field
\cite{phant} and f-essence \cite{f}. Another interesting set of
proposals to DE puzzle is the `modified gravity' (including $F(T)$, $F(R)$, $F(G),$ etc)  which was proposed
after the failure of general relativity to explain the DE puzzle. This new set of
gravity theories passes several solar system and astrophysical tests
\cite{sergei2}.


A gravitational theory can be constructed on a non-Riemannian (
Weitzenbock) manifold where the properties of gravity are determined
through torsion of spacetime and not curvature. Some earlier
attempts in this direction were made by Einstein himself and other
researchers. A recent version of torsion based gravity is $F(T)$
\cite{f(T)}, where $T$ is the torsion scalar constructed from the
tetrad.  Choosing $F(T)=T$, leads to the teleparallel gravity
\cite{hayashi,hehl} and is in good agreement with some standard
tests of the general relativity in solar system \cite{hayashi}.
Numerous features of theoretical interest have been studied in this
gravity already including Birkhoff's theorem \cite{birkhoff},
cosmological perturbations \cite{zheng} and phantom crossing of the
state parameter \cite{bamba}. Moreover, the local Lorentz invariance
is violated which henceforth leads to violation of first law of
thermodynamics \cite{miaoli,sotirio}. Also the entropy-area relation
in this gravity takes a modified form \cite{bomb}. The Hamiltonian
structure of $F(T)$ gravity has been investigated  and found that
there are five degrees of freedom \cite{miao}. The torsion based
theory is  also an alternative candidate to the mechanism of cosmic
inflation \cite{inflation}.


In teleparallel gravity, the equations of motion for any geometry
are exactly the same as of general relativity. Due to this reason,
the teleparallel gravity is termed as `teleparallel equivalent of
general relativity'. In teleparallel gravity, the dark energy puzzle
is studied by introducing a scalar field with a potential. If this
field is minimally coupled with torsion, then this effectively
describes quintessence dark energy. However if it is non-minimally
coupled with torsion, than more rich dynamics of the field appears
in the form of either quintessence or phantom like, or by
experiencing a phantom crossing \cite{geng}. Xu et al \cite{xu}
investigated the dynamics and stability of a canonical scalar field
non-minimally coupled to gravity (arising from torsion). They found
that the dynamical system has an attractor solution and rich
dynamical behavior was found. In the context of general relativity,
a scalar field non-minimally coupled with gravity has been studied
in \cite{mark}. We here extend these previous studies by replacing
$T$ with an arbitrary function $F(T)$. We found that such a
dynamical system possess no stable equilibrium point, however rich
dynamical behavior of quintessence and phantom energy is observed.


We follow the following plan: In section II we write the action  and
equations of motion of our model. In section III, we give motivation
for choosing particular forms of model functions. In section IV, we
write the dynamical system in dimensionless form and discuss its
stability and phase space behavior. In section V, we discuss the
cosmological implications of our model related to present
accelerated Universe. We provide conclusion in section VI.

\section{Basic equations}

We are interested in conformally invariant models i.e. models which remain invariant under conformal transformation. Under the rules $e^i_\mu\rightarrow\Omega(x)e^i_\mu$ (or $g_{\mu\nu}\rightarrow \Omega(x)g_{\mu\nu}$), $\phi\rightarrow\Omega(x)^{-1}\phi$, the equations of motion and the whole action remain invariant of a conformally invariant model. Here $e^{i}_{\mu}$ is the tetrad (vierbein) basis. In general for a $D-$dimensional gravity model, the conformal coupling parameter $\xi=\frac{D-2}{4(D-1)},$ so that  in a four dimensional theory, $\xi=\frac{1}{6}$ \cite{gv}.
Further $F(T)$ theories from dynamical point of view are completely different under conformal transformation unlike $f(R)$ theory. Its not possible to rewrite the total action of $F(T)$ gravity in form of teleparallel action plus scalar field \cite{epl}. It shows that even the pure $F(T)$ model behaves differently under conformal transformation. We propose an action of $F(T)$ gravity conformally and non-minimally coupled with a scalar field as\footnote{Here $8\pi G=1$}
\begin{eqnarray}\label{action}
\mathcal{S}=\int d^4x~ e\Big[\frac{F(T)}{2}(1+\xi \phi^2)+\frac{1}{2}\epsilon \dot \phi^2-V(\phi)+\mathcal{L}_m\Big],
\end{eqnarray}
where $e=\text{det}(e^{i}_{\mu})$. Here $\xi$ is a conformal
coupling parameter of order unity while $\epsilon=+1,-1$ represents
quintessence and phantom energy respectively. The dynamical quantity
of the model is the tetrad and scalar field $\phi$ with a scalar
potential $V(\phi)$. $\mathcal{L}_m$ is the matter Lagrangian. It is
assumed that both matter and scalar field are distributed as perfect
fluid. The action (\ref{action}) can be considered as a
generalization of a teleparallel gravity non-minimally coupled with
a scalar field \cite{xu}. We add some more comments on our model:
the minimally coupled scalar fields are not conformal invariant,
otherwise $F(T)$ gravity is not local Lorentz invariance. After the
Lorentz symmetry breaking, we prefer our model must have conformal
symmetry. Just for pure $F(T)$, it is well-known that it can be
possible to write the pure $F(T)$ action in a conformal gauge, like
$F(R)$. If we couple matter (here scalar field) with $F(T)$, this
matter part must have the same (conformal) symmetry which can be
interpreted as a generalized conformal invariance.

We assume a spatially flat Friedmann-Robertson-Walker (FRW) metric as a background spacetime
\begin{equation}\label{frw}
ds^2=-dt^2+a(t)^2[dx^2+dy^2+dz^2],
\end{equation}
where $a(t)$ is a scale factor and $e^{i}_{\mu}=(1,a(t),a(t),a(t))$. The equations of motion are obtained by varying the action (\ref{action}) w.r.t. $a(t)$ and $\phi(t)$, we get
\begin{equation}
6H^2F_T(1+\xi\phi^2)+\frac{1}{2}(1+\xi\phi^2)F=\rho_{m}+\frac{1}{2}\epsilon\dot\phi^2+V(\phi),\label{f1}
\end{equation}
\begin{equation}
\ddot\phi+3H\dot\phi-\epsilon[\xi\phi F-V'(\phi)]=0,\label{f2}
\end{equation}
which are Friedmann and Klein-Gordon equations respectively.

The second Friedmann equation is
\begin{eqnarray}\label{ddot}
\frac{\ddot a}{a}&=&\frac{1}{4a(1+\xi\phi^2)(-12F_{TT}\dot a^2+a^2F_T)}\Big[ -8a\dot a F_T((1+\xi\phi^2)+\xi a\phi\dot\phi) \nonumber\\&&-48(1+\xi\phi^2)\dot a^4 a^{-1}F_{TT} -a^3((1+\xi\phi^2)F+\epsilon\dot\phi^2-2V(\phi)) \Big].\label{f3}
\end{eqnarray}
We can rewrite (\ref{f1}) as
\begin{equation}\label{1}
3H^2=\rho_\phi^{\text{eff}}+\rho_m,
\end{equation}
or
\begin{equation}\label{1}
1=\Omega_\phi^{\text{eff}}+\Omega_m,\ \ \ \Omega_\phi^{\text{eff}}\equiv\frac{\rho_\phi^{\text{eff}}}{3H^2},\ \Omega_m\equiv\frac{\rho_m}{3H^2},
\end{equation}
where $\rho_\phi^{\text{eff}}$ is the effective energy density of scalar field written as
\begin{eqnarray}
\rho^{\text{eff}}_\phi &\equiv& \frac{1}{2}\epsilon\dot\phi^2+V(\phi)-\frac{T}{2}+\frac{(1+\xi\phi^2)}{2}(2TF_T-F).
\end{eqnarray}
Combining Eqs. (\ref{ddot}) and (\ref{1}),we get
\begin{equation}\label{Hdot}
2\frac{\ddot a}{a}+H^2=-p^{\text{eff}}_\phi,
\end{equation}
where $p^{\text{eff}}_\phi$ is the effective pressure of scalar field given by
\begin{eqnarray}
p^{\text{eff}}_\phi&\equiv&\frac{T}{6}-\frac{1}{2a(1+\xi\phi^2)(-12F_{TT}\dot a^2+a^2F_T)}\Big[ -8a\dot a F_T((1+\xi\phi^2)+\xi a\phi\dot\phi)\nonumber\\&&-48(1+\xi\phi^2)\dot a^4 a^{-1}F_{TT}  -a^3((1+\xi\phi^2)F+\epsilon\dot\phi^2-2V(\phi)) \Big].
\end{eqnarray}
The deceleration parameter for this model is
\begin{eqnarray}\label{ddot1}
q=-\frac{\ddot a}{aH^2}&=&\frac{-1}{4a(1+\xi\phi^2)(-12F_{TT}\dot a^2+a^2F_T)H^2}\Big[ -8a\dot a F_T((1+\xi\phi^2)+\xi a\phi\dot\phi) \nonumber\\&&-48(1+\xi\phi^2)\dot a^4 a^{-1}F_{TT} -a^3((1+\xi\phi^2)F+\epsilon\dot\phi^2-2V(\phi)) \Big].
\end{eqnarray}

\begin{figure*}[thbp]
\begin{tabular}{rl}
\includegraphics[width=7.5cm]{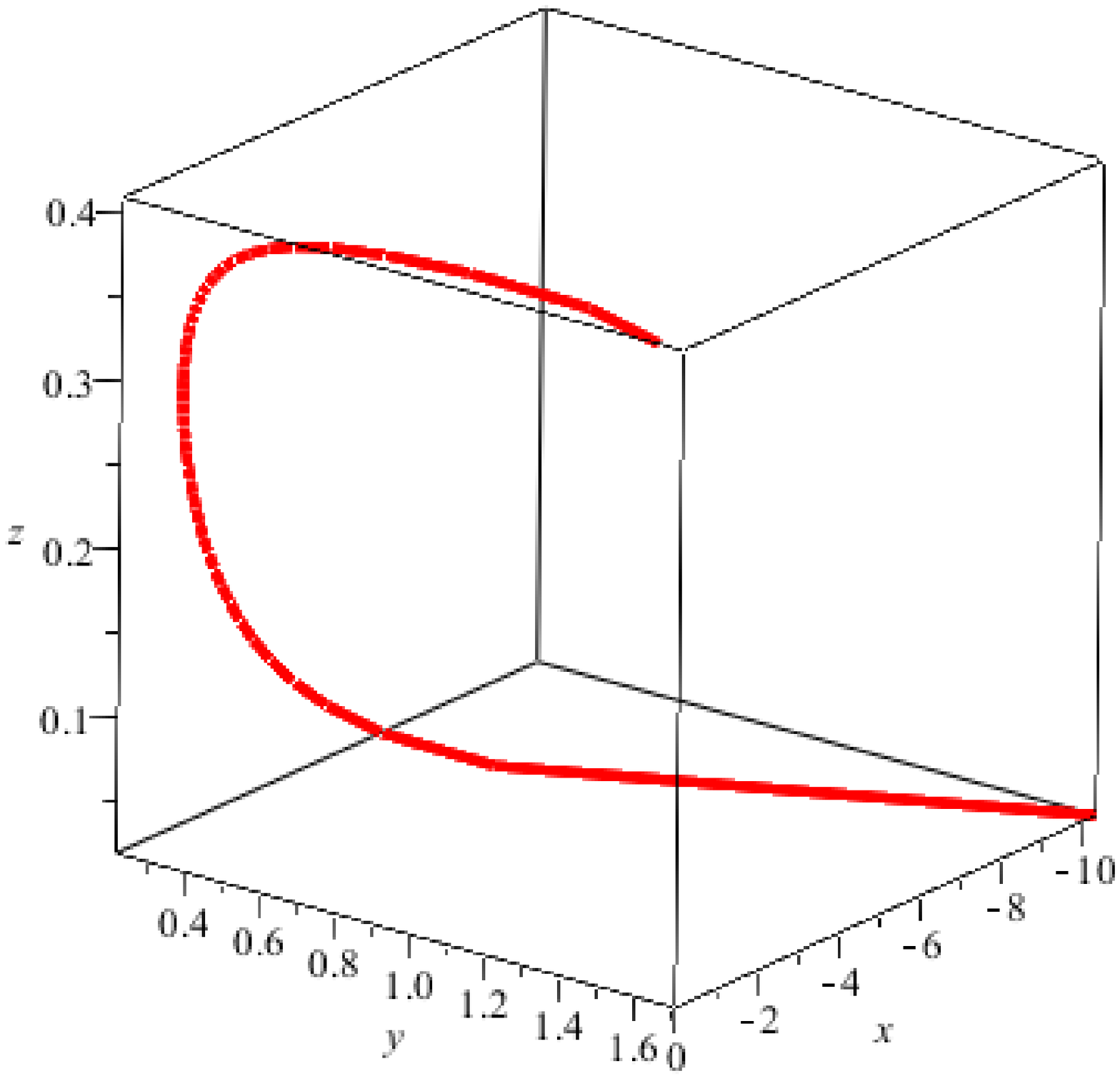}&
\includegraphics[width=7.5cm]{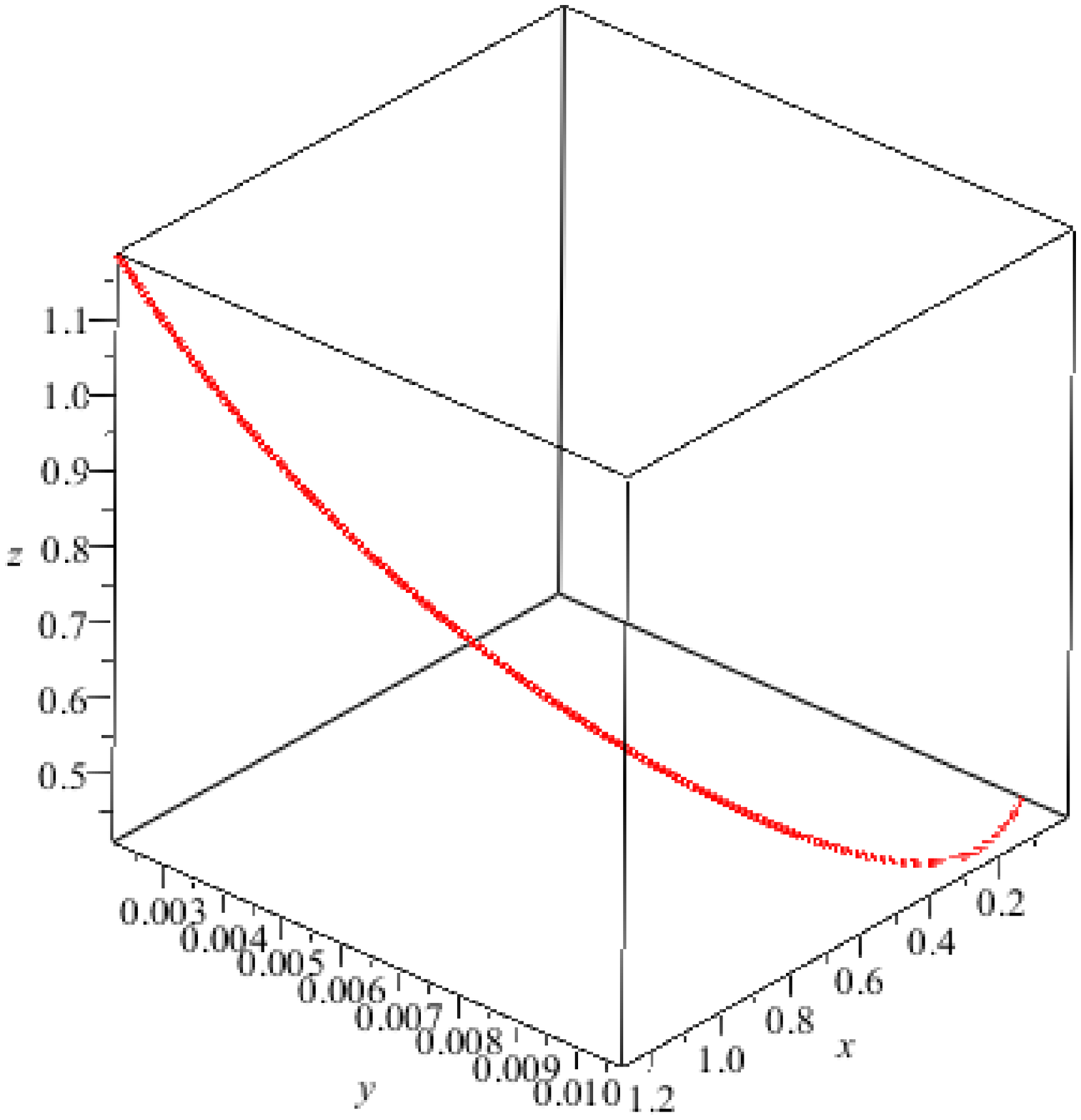} \\
\includegraphics[width=7cm]{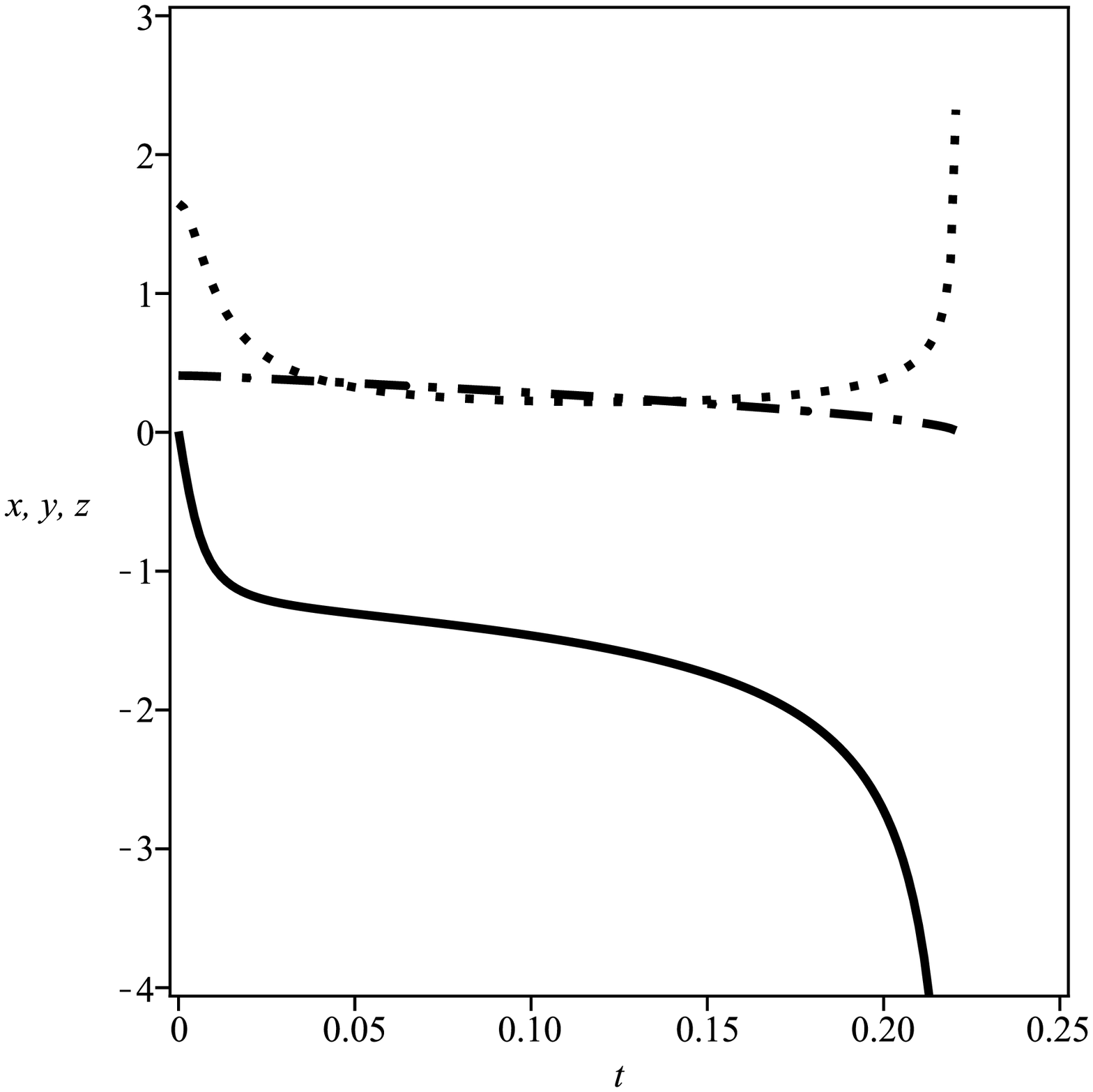}&
\includegraphics[width=7cm]{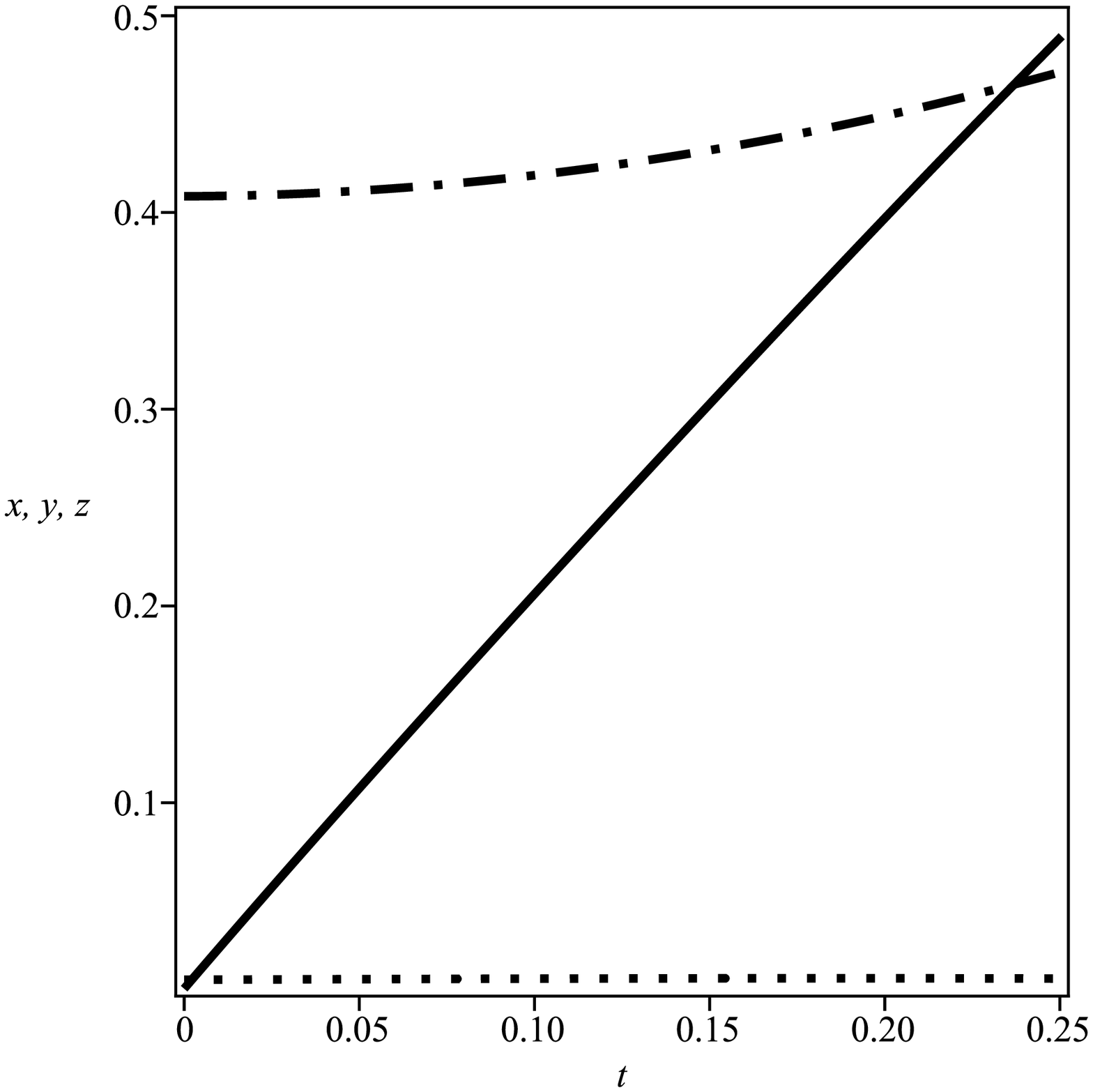} \\
\end{tabular}
\caption{ (\textit{Top Left}) Three dimensional phase portrait of the dynamical system with $\epsilon=+1$.   (\textit{Top Right}) Phase space diagram of the dynamical system with $\epsilon=-1$. (\textit{Bottom Left}) Time evolution of dynamical parameters for $\epsilon=+1$. (\textit{Bottom Right}) Time evolution of dynamical parameters for $\epsilon=-1$. In the lower two figures, $x$, $y$, $z$ are represented by line, dot, dot-dash respectively.}
\end{figure*}

\begin{figure*}[thbp]
\begin{tabular}{rl}
\includegraphics[width=7cm]{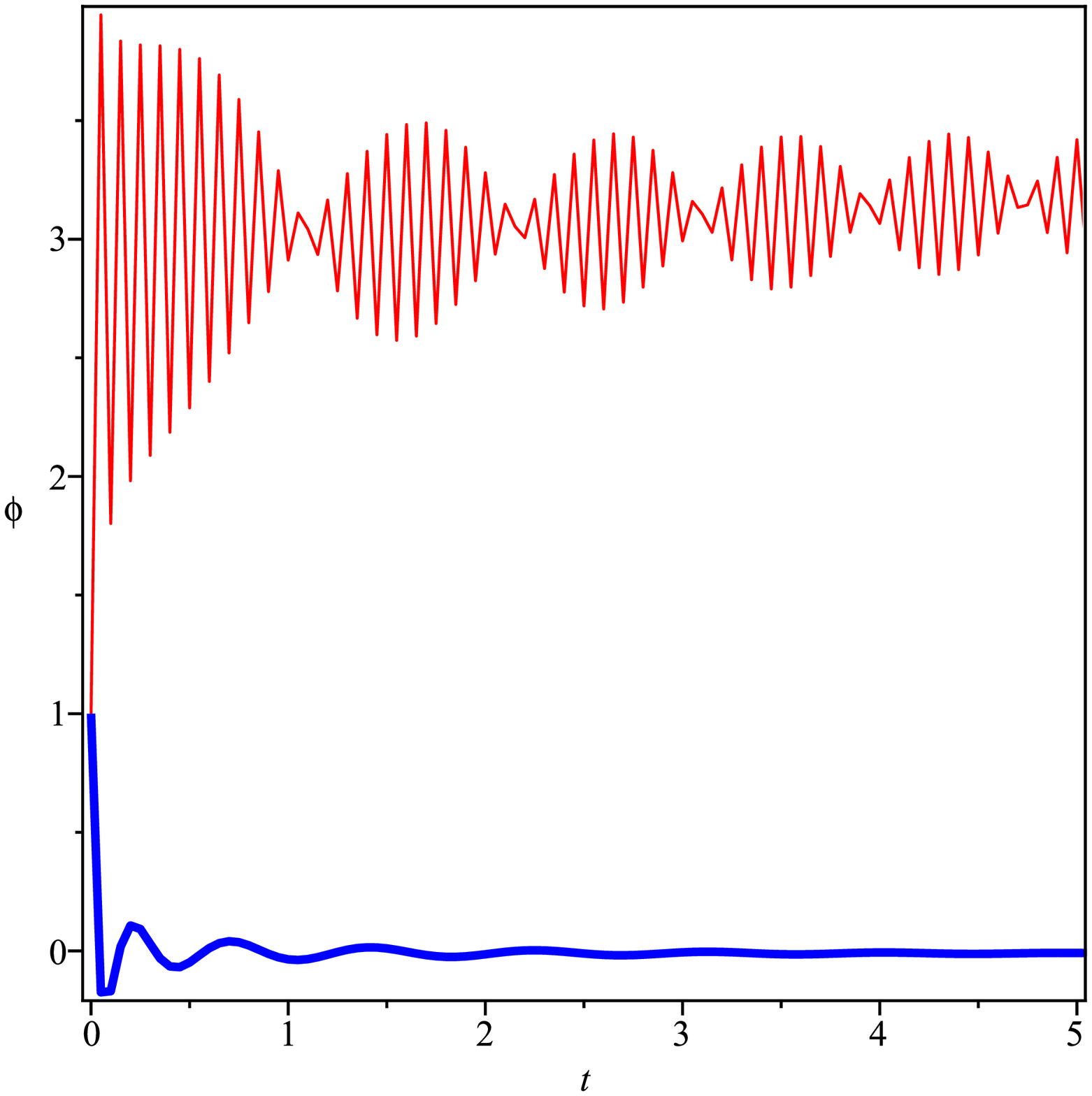}&
\includegraphics[width=7cm]{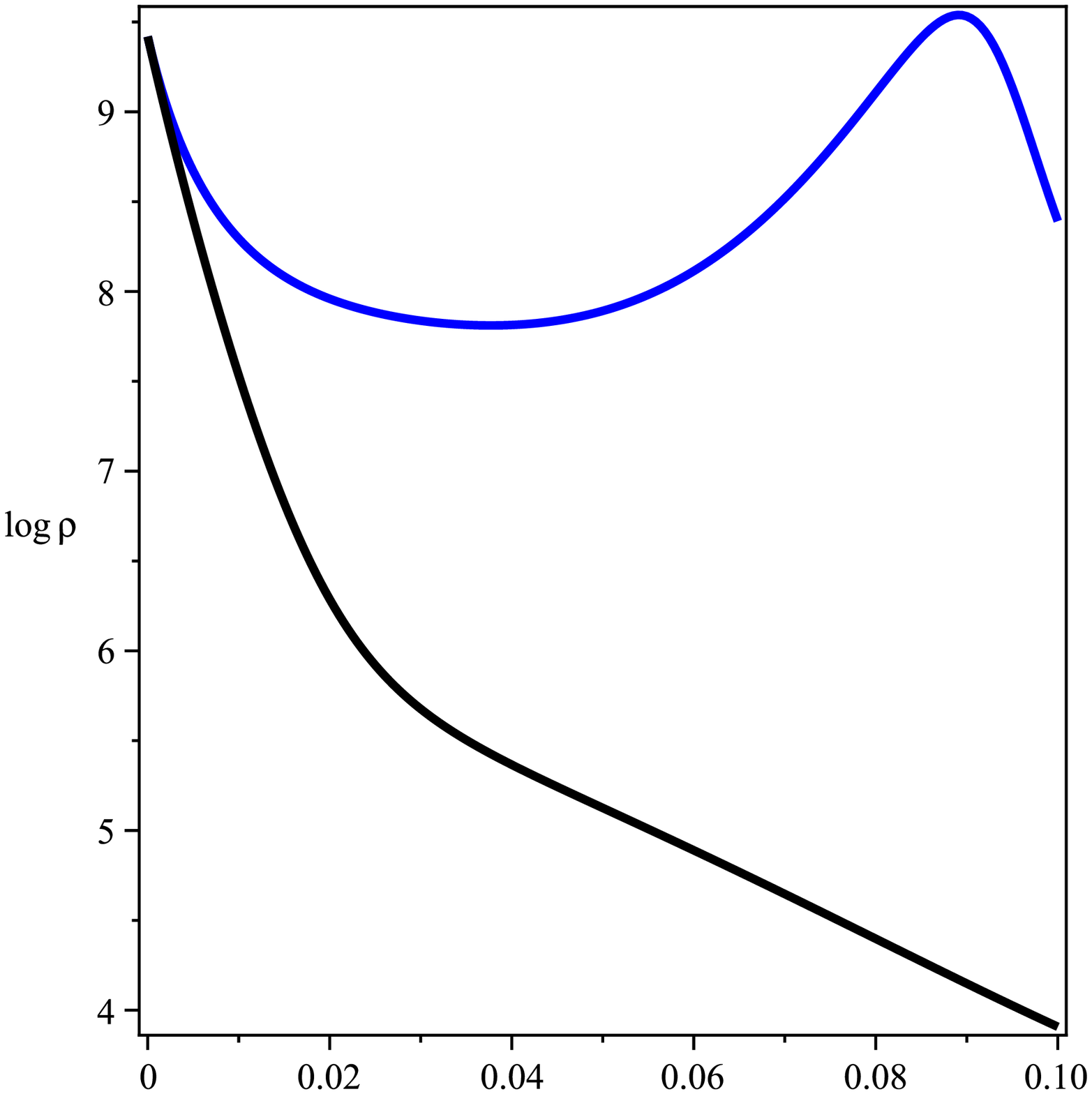} \\
\includegraphics[width=7cm]{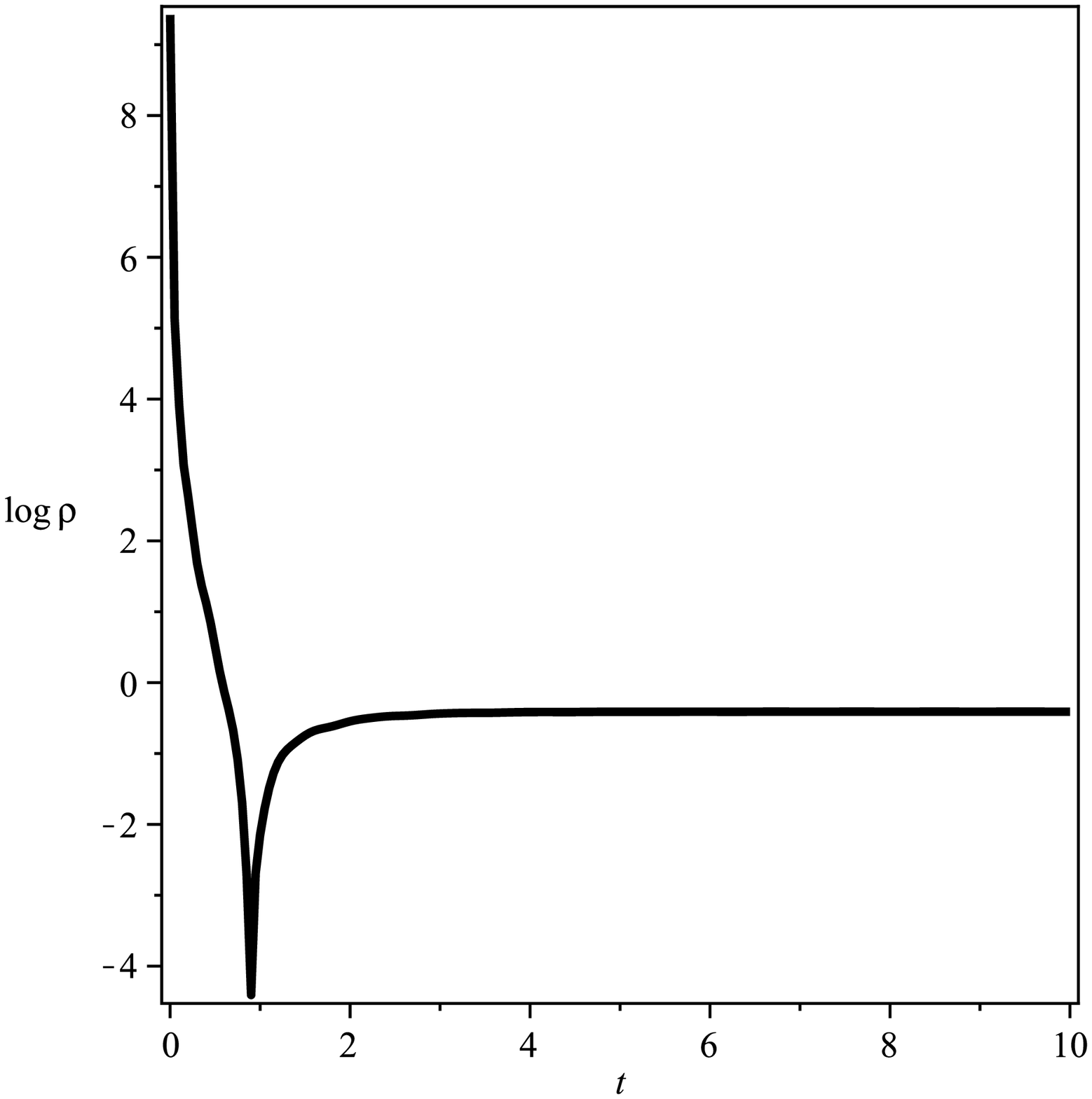}&
\includegraphics[width=7cm]{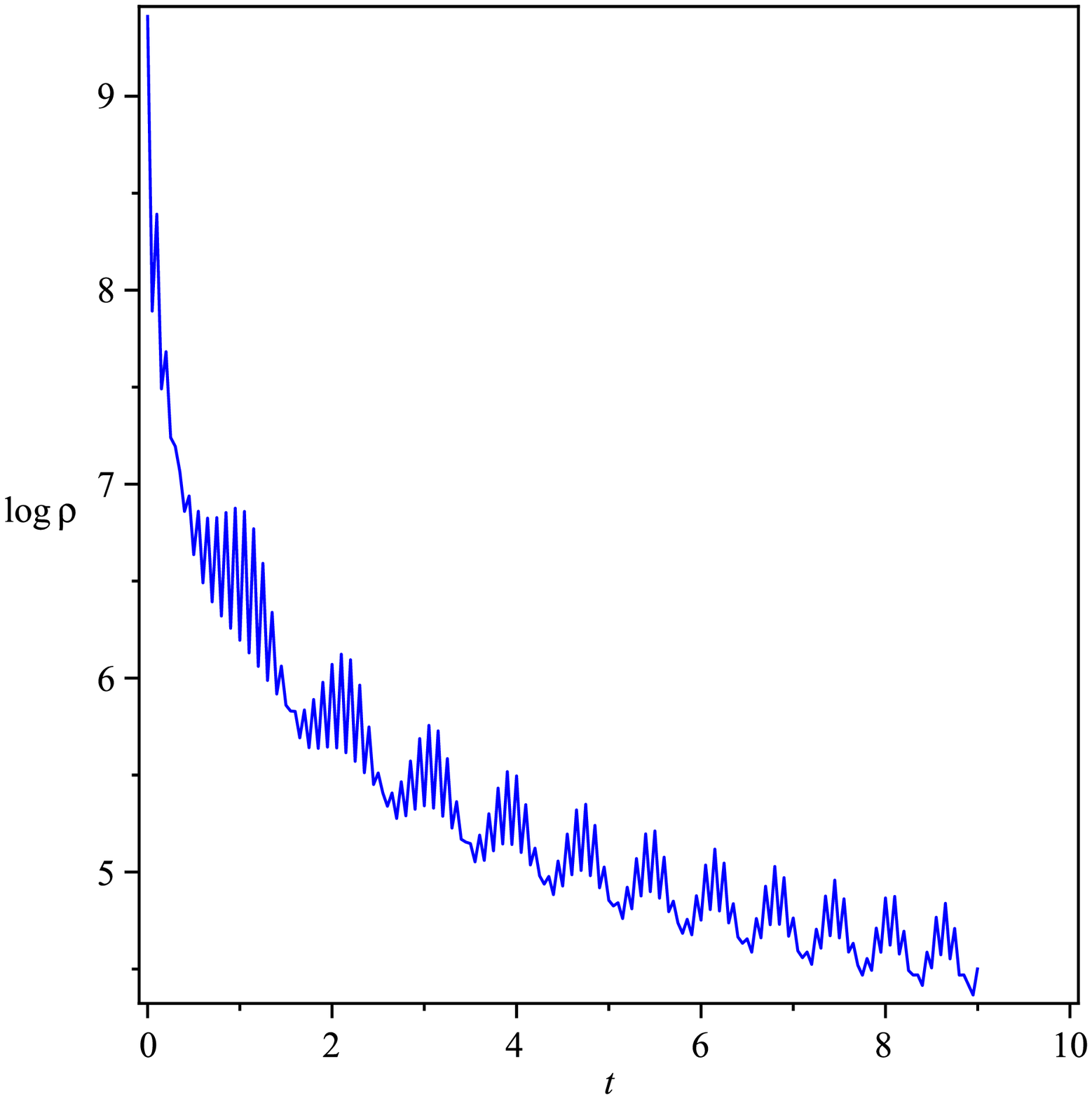} \\
\end{tabular}
\caption{(\textit{Top Left}) Time evolution of scalar fields: phantom energy (red) and quintessence (blue).    (\textit{Top Right}) Time evolution of energy densities for small time: phantom energy (blue) and quintessence (black).  (\textit{Bottom Left}) Time evolution of energy density of quintessence field for large times.  (\textit{Bottom Right}) Time evolution of energy density of phantom energy for large times. }
\end{figure*}

\section{Choice of $F(T)$ and $V(\phi)$ }

We pick a suitable
$f(T)$ expression which contains a constant, linear and a non-linear
form of torsion, specifically \cite{at}
\begin{equation}\label{ft}
F(T)=2c\sqrt{-T} +\alpha T+C_2,
\end{equation}
where $\alpha$, $c$ and $C_2$ are arbitrary constants \footnote{Here
$c$ does not represent the speed of light}. The first and the third
terms (excluding the middle term) has correspondence with the
cosmological constant EoS in $f(T)$ gravity \cite{mirza}. There are
many kinds of such these models, reconstructed from different kinds
of the dark energy models. For example this form (\ref{ft}) may be
inspired from a model for dark energy from proposed form of the
Veneziano ghost \cite{kk}. Recently Capozziello et al \cite{capo123}
investigated the cosmography of $F(T)$ cosmology by using data of
BAO, Supernovae Ia and WMAP. Following their interesting results, we
notice that if we choose $2c\equiv\sqrt{6}H_0(\Omega_{m0}-1)$, than
one can estimate the parameters of this $F(T)$ model as a function
of Hubble parameter $H_0$ and the cosmographic parameters and the
value of matter density parameter. It is interesting to note that
reconstruction of $F(T)$ model according to holographic dark energy
\cite{hde} leads to the same model as (\ref{ft}).

Concerning the scalar potential, we choose an exponential function which has numerous implications in cosmological inflation \cite{in} and dark energy in the present Universe \cite{de}
\begin{equation}\label{V}
V(\phi)=V_0e^{\beta\phi},
\end{equation}
where $\beta$ and $V_0$ are constants. Using (\ref{V}), it has been
shown in literature \cite{de} that transition of dark energy state
parameter across the cosmological boundary is possible. Also this
exponential potential is useful in assisting inflation. Without loss
of generality, we assume $\beta>0$. But here we focus only on
late-time evolution of the scalar field and we do not give any
prediction of our model for inflation. In the limit
$\beta\rightarrow0$, we recover the constant potential case and
therefore the model is continuously connected with $\Lambda$CDM. As
a remark,  our choices for scalar potential and the $F(T)$ can
independently be responsible for DE, but in the present context of
non-minimal coupling, both scalar field and $F(T)$ couples
non-minimally to generate the desired effect of cosmic acceleration.

\section{Analysis of stability in phase space}

We define dimensionless density parameters via
\begin{equation}\label{3}
x\equiv  \frac{\dot\phi}{\sqrt{6}H}  ,\ \
y\equiv  \frac{\sqrt{V}}{\sqrt{3}H} ,\ \
z\equiv   \sqrt{\xi}\phi .
\end{equation}
Here $x^2$ and $y^2$ represent the density parameters of the
kinetic and potential terms respectively. We expect interesting
cases to have the scalar field rolling down the
slope of the potential, as $\beta > 0$,
we should have $x > 0$.
The  equations  in dimensionless variables
reduce to
\begin{eqnarray}\label{sys1}
\frac{dx}{dN}&=&-2x-{\frac {\sqrt {6{\xi}^{-
1}}\alpha\,z}{\epsilon}}-{\frac {\sqrt {6\xi}z}{\epsilon}}\nonumber\\&&+2\,
\xi\,c\sqrt {3{\xi}^{-1}}{\it yz}\,{{\rm e}^{1/2\,{\frac {
\beta\,z}{\sqrt {\xi}}}}}{\epsilon}^{-1}\\&&-\frac{1}{2}\,{\frac {\beta\,\sqrt {6}
{y}^{2}}{\epsilon}}-x  ( -\alpha\, ( 1+1/2
\,{z}^{2} ) -\frac{1}{2}-{z}^{2}\nonumber\\&&-2\,\xi\,\sqrt {6}\sqrt {{\xi}^{-1}}
 ( \alpha+1 ) x  z-6\,\epsilon\,  x
  ^{2}+\frac{3}{2}\,{y}^{2}\nonumber\\&&+2\,\xi\,cx \sqrt {3{\xi}^{-1}}yz{{\rm e}^{1/2\,{\frac {\beta\,z}
{\sqrt {\xi}}}}} )  ( \alpha+1 ) ^{-1} ( 1+{z}^{
2} ) ^{-1}
,\nonumber\\
\frac{dy}{dN}&=&-\frac{3}{2}\,\beta\,\sqrt {6}yx-y ( -\alpha
\, ( 1+1/2\,{z}^{2} ) -\frac{1}{2}-{z}^{2}\nonumber\\&&-2\,\xi\,\sqrt {6
{\xi}^{-1}} ( \alpha+1 ) xz-6\,\epsilon\,{x}^{2}+\frac{3}{2}\,{y}^{
2} \\&&+2\,\xi\,cx\sqrt {3{\xi}^{-1}}yz{{\rm e}^{1/2\,{\frac {\beta
\,z}{\sqrt {\xi}}}}} )  ( \alpha+1 ) ^{-1} ( 1+{
z}^{2} ) ^{-1},\label{sys2}
\nonumber\\
\frac{dz}{dN}&=&\sqrt {6\xi}x,\label{sys3}
\end{eqnarray}
where $N\equiv\ln a$, is called the e-folding parameter.
To discuss the stability of the system (\ref{sys1})-(\ref{sys3}), we first obtain the critical points by solving the equations ($\frac{dx}{dN}=0,\frac{dy}{dN}=0,\frac{dz}{dN}=0$). We linearize the system near the critical points up to first order. After constructing a jacobian matrix of coefficients of linearized system, we find its eigenvalues. If all eigenvalues are negative, the corresponding critical point is stable (attractor), otherwise an unstable point.

  \begin{table}[ht]

  \centering
  \begin{tabular}{|c| c |c| c|c |c|}
  \hline\hline
    Point & $(x_*,y_*,z_*)$ & $\lambda_1$ & $\lambda_2$ &  $\lambda_3$ &Stability  Condition \\ [0.5ex]
     \hline
      A & $(0,0,0)$ &$\frac{5+4\alpha}{2(1+\alpha)}$ &
$\frac{2\alpha+1}{2(1+\alpha)}$ & 0&Unstable \\\hline
      B & $(0,0,z_*)$ &$\frac{z_*^2(4-5\xi)+(5-4\xi)}{2(1+z_*^2)(\xi-1)}$ & $\frac{z_*^2(\xi-2)+2\xi-1}{2(1+z_*^2)(\xi-1)}$ &0& Unstable\\
      \hline
      \end{tabular}
        \caption{Critical points and stability conditions.}
       \label{table:fluid}
       \end{table}

 In Table I, we present the critical points, corresponding eigenvalues and stability condition. We notice that the system of dynamical equations admit no stable critical point. The first critical point A is trivial one while point B contains an undetermined component $z_*$. For A, $\lambda_{1,2}>0,$ $ \lambda_3=0$ while for B, $\lambda_1<0$, $\lambda_2>0$, $\lambda_3=0$, thus both A and B are unstable points.

\begin{figure}
\centering
 \includegraphics[scale=0.4] {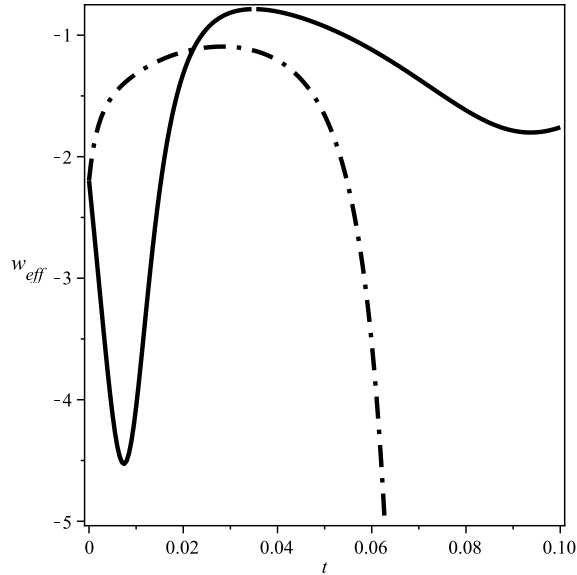}
  \caption{ Time evolution of effective EoS parameter for scalar field $w_{\text{eff}}$. Solid line (dot-dash) corresponds to $\epsilon=+1$ ($\epsilon=-1$).    }

\end{figure}

\section{Cosmological Implications}

In this section, we will give some cosmological implications of our model by numerically solving the dynamical equations. In the top panel of figure-1, the three dimensional phase space of $x$, $y$, $z$ is plotted for two different values of $\epsilon$. The top left figure (for $\epsilon=-1$) shows that the trajectory starts from $x=-10<0$ causing a negative friction term $-3H\dot\phi$ in the Klein-Gordon equation. The friction term gets less dominant while scalar field and potential energy parameters dominate in the later evolution. It shows that the late stage evolution is determined by scalar potential alone. This fact is also evident from bottom left figure that only potential term dominates the dynamics while scalar field and the kinetic term do not contribute in cosmological dynamics. However, putting $\epsilon=+1$ in the dynamical system reverses the dynamical evolution of system as shown in right panel in Fig.1. The scalar potential remains vanishing while the cosmic dynamics are determined by kinetic term $\dot\phi$ and scalar field. It shows that  $\epsilon=+1(-1)$ leads to kinetic term (scalar potential) dominated regimes in the late time evolution, despite the system evolves from same initial conditions ($a(0)=1$, $\dot a(0)=H_0=74.2$, $\phi(0)=1$, $\dot\phi(0)=1$, $\xi=1/6$).

In Figure 2 we show the time evolution of scalar fields and logarithmic energy densities for $\epsilon=\pm1$. The top left figure shows that for phantom energy, the field undergoes successive stages of fluctuations after a time gap of unity. Here the amplitude of phantom scalar field in each successive stage becomes progressively less than the previous stage. In the late time evolution, it is expected that the field will lose its energy and fluctuations decrease, as seen in bottom right figure. However for quintessence, the top left figure shows that field decays for $t<0.5$, and corresponding log energy density stays around $-1$ ($\rho\sim e^{-1}$).

We define the equation of state parameter of the scalar field
\begin{equation}
w_{\text{eff}}\equiv\frac{p^{\text{eff}}_\phi}{\rho_\phi^{\text{eff}}},
\end{equation}
where
\begin{eqnarray}\label{111a}
\rho_\phi^{\text{eff}}&=&\frac{1}{2}\epsilon\dot\phi^2+V(\phi)-3H^2[(1+\alpha)\xi\phi^2+\alpha],\\
p^{\text{eff}}_\phi&=&\frac{-2}{(1+\alpha)(1+\xi\phi^2)}\Big[ -\frac{(\alpha+1)}{2}H^2(1+\xi\phi^2)\nonumber\\&&-2\xi(1+\alpha)\phi\dot\phi H-\frac{\epsilon}{4}\dot\phi^2+\frac{1}{2}V(\phi)+\frac{c\sqrt{6}\xi}{3}\phi\dot\phi\Big]\nonumber\\&&-\frac{1}{3}\Big[ \rho_m+\frac{1}{2}\epsilon\dot\phi^2+V(\phi)-3H^2[(\alpha+1)\xi\phi^2+\alpha]\Big ].\label{pdot}
\end{eqnarray}
In figure-3, we plot the time evolution of the effective EoS parameter of the scalar field $w_{\text{eff}}$ for two available values of $\epsilon$. Adopting the same initial conditions as in previous figures, we observe that the state parameters for both models evolve from the same initial value $w_{\text{eff}}(0)\simeq-2$ which is a phantom state for the state parameter. However the later evolution follows opposite trajectory i.e. for $\epsilon=-1$, the state parameter goes on to take more negative values with time, thereby evolving to super-phantom state. For $\epsilon=+1$, the state parameter initially behaves like phantom and slowly evolves to cross the $w_{\text{eff}}=-1$ boundary, while again returning to phantom regime in the late time evolution.

\section{Conclusion}

To summarize, we investigated the stability  and phase space of a
non-minimally  conformally coupled scalar field  in $F(T)$
cosmology.  We found that the dynamical system of equations admit
two unstable critical points, thus no attractor solutions exist in
this cosmology. Treating the scalar field as quintessence and
phantom energy separately, we found that for phantom energy, the
late stage evolution is determined by the potential energy in phase
space while for quintessence case, it is kinetic energy term that
plays central role. Furthermore, the time evolution of phantom
scalar field undergoes progressive stages of fluctuations while
loosing energy density and ending up to the value $e^4$ in the
relevant scale. For quintessence, the scalar field remains stable
for long time with a constant energy density.  It is interesting to
note that our model correctly predicts the present state of the
Universe is dominated by phantom energy while it will remain so in
the far future. Further, we observed that the transition of the
state parameter crosses the cosmological boundary twice in the case
of quintessence field only. Also we noted that $\epsilon=+1(-1)$
leads to kinetic term (scalar potential) dominated regimes in the
late time evolution, despite the system evolves from same initial
conditions.

\end{document}